\documentclass[epsfig]{elsart} 
\usepackage{graphics}
\begin{document}
\begin{frontmatter}
\begin{flushright}
\vglue-2.0cm
{\small\bf Invited talk, X. International Light-Cone\\
 Meeting, HD2000, Heidelberg, June 2000}
\end{flushright}
\title{Measurement of Light-Cone Wave Functions by Diffractive Dissociation }
\author{Daniel Ashery} \footnote{
Supported in part by the Israel Science
Foundation and the US-Israel Binational Science Foundation.}
\address{School of Physics and Astronomy, Raymond and Beverly Sackler
Faculty of Exact Sciences Tel Aviv University, Israel}

\date{2 August 2000 }
\begin{abstract}
Diffractive dissociation of particles can be used to study their
light-cone wave function. Results from Fermilab experiment E791 for
diffractive dissociation of 500 GeV/c $\pi^-$ mesons into di-jets 
are presented. The results show that the $|q\bar
{q}\rangle $ light-cone asymptotic wave function 
describes the data well for $Q^2 \sim 10 ~{\rm (GeV/c)^2}$ or more. 
Evidence for color transparency comes from a measurement of
the $A$-dependence of the yield of the diffractive di-jets. It is
proposed to carry out similar studies for the light-cone wave function
of the photon.
\end{abstract}
\end{frontmatter}

\section*{Introduction}
\subsection*{The Pion Light-Cone Wave Function}
The internal momentum distribution of valence quarks in hadrons
are fundamental to QCD. \cite{stan1}. They are generated from
the valence light-cone wave functions integrated over $k_t < Q^2$, where $k_t$
is the intrinsic transverse momentum of the valence constituents and
$Q^2$, the total momentum transfer squared. Even though these amplitudes
were calculated about 20 years ago, there have been no direct measurements
until those reported here. 

\begin{figure}
\begin{center}
\resizebox{1.0\textwidth}{!}{
 \includegraphics{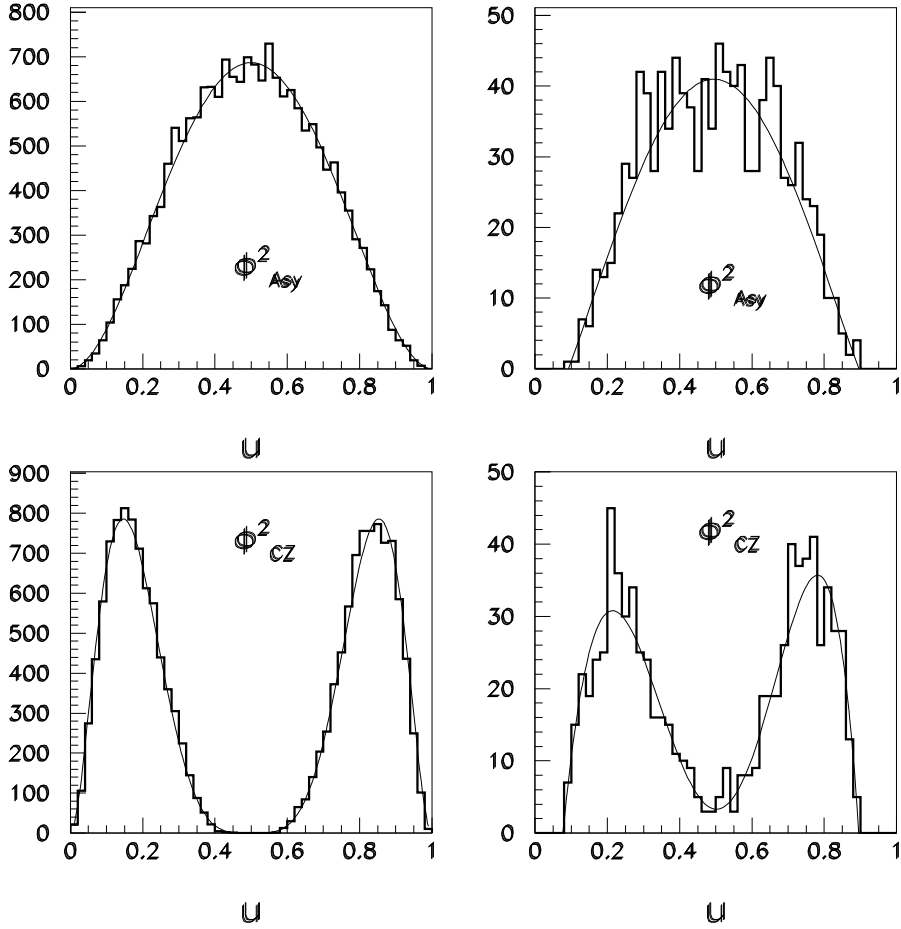}
}\caption{Monte Carlo simulations of squares of the two wave functions at the
quark  level (left) and of the reconstructed distributions of di-jets as
detected  (right).  $\phi_{Asy}^2$ is the asymptotic function (squared) and
$\phi_{CZ}^2$ is the Chernyak-Zhitnitsky  function (squared).
The di-jet mass used in the simulation is 6 GeV/c$^2$
and the plots are for $1.5 ~\rm{GeV/c} ~\leq ~k_t ~\leq ~2.5
~\rm{GeV/c}$.
\vspace*{5 mm}
\label{fig:x_mc}
}\end{center}
\end{figure}

The pion wave function can be expanded in terms of Fock states:
\begin{equation}
\Psi = a_1 |q\bar {q}\rangle + a_2 |q\bar {q}g\rangle +
           a_3|q\bar {q}gg\rangle + ... .
\end{equation}
Two functions have been proposed to describe the momentum distribution
amplitude for the quark and antiquark in the $|q\bar {q}\rangle$
configuration. The asymptotic function was calculated  using perturbative QCD
(pQCD) methods \cite{lb,er,bbgg}, and is the solution to the pQCD
evolution equation for very large $Q^2$ ($Q^2 \rightarrow \infty$):
\begin{equation}
\phi_{as}(u) =\sqrt{3} u(1-u).
\label{asy}
\end{equation}
$u$ is the fraction of the longitudinal momentum of the pion
carried by the quark in the infinite momentum frame. It should not be 
confused with $x_{Bj}$ which is not specific to a single Fock state.
The antiquark carries
a  fraction ($1 - u$). Using QCD sum rules, Chernyak and Zhitnitsky
\cite{cz} proposed a function that is expected to be correct for low
Q$^2$:
\begin{equation}
\phi_{cz}(u) =5\sqrt{3} u(1-u)(1-2u)^2.
\label{cz}
\end{equation}
As can be seen from eqns. \ref{asy} and \ref{cz} and from Fig.
\ref{fig:x_mc}, there is a large difference between the two functions.
Measurements of form factors are insensitive to the wave function as
these quantities are derived by integrating over the wave function 
and interpretation of the results is model dependent \cite{stesto}.
In this work we describe an experimental study that  maps the momentum 
distribution of the valence 
$|q\bar {q}\rangle$ in the pion. This provides the first direct
measurement of the pion light-cone wave function (squared).
The concept of the measurement is the following: a high energy pion
dissociates diffractively on a heavy nuclear target. 
This is a coherent process in which the quark and antiquark break 
apart and hadronize
into two jets. If in this fragmentation process the quark momentum
is transferred to the jet, measurement of the jet momentum gives the
quark (and antiquark) momentum. Thus:
$u_{measured} = \frac {p_{jet1}} {p_{jet1}+p_{jet2}}.$
From simple kinematics and assuming that the masses
of the jets are small compared with the mass of the di-jets, the virtuality
and mass-squared of the di-jets are given by:
$Q^2 \sim  M_{DJ}^2 = \frac{k_t^2}{u(1 - u)}$ where $k_t$ is the transverse 
momentum of each jet. By studying the momentum
distribution for various $k_t$ bins, one can observe changes in the apparent
fractions of asymptotic and Chernyak-Zhitnitsky (CZ) contributions to the pion
wave function.

The basic assumption that the momentum carried by the dissociating
$q \bar q$ is transferred to the di-jets was examined by Monte Carlo
(MC) simulations of the asymptotic and (CZ) wave functions (squared).
The MC samples were allowed to hadronize through the LUND PYTHIA-JETSET
model \cite{mc} and then passed through simulation of the experimental
apparatus (described in the next section) to simulate the effect of 
unmeasured neutrals and other experimental distortions.
In Fig. \ref{fig:x_mc} the initial distributions at the
quark level are compared with the final distributions of the detected
di-jets. As can be seen, the qualitative
features of the distributions are retained. The results of this analysis
come from comparing the observed $u$-distribution to a combination of the
distributions shown, as examples, on the right of Figure \ref{fig:x_mc}.


\subsection*{The Color Transparency Effect}
The Color-Transparency (CT) phenomenon is derived from the prediction that
color fields cancel for physically small color singlet systems of quarks
and gluons \cite{lonus}. This effect of color neutrality (or color screening)
is expected to lead to the suppression of initial and final state
interactions for a small sized system or point-like configuration (PLC)
formed in a large angle hard process \cite{ct}. Observation
of CT requires that a PLC is formed and that the energies are
high enough so that expansion of the PLC does not occur \cite{bm}
(the frozen approximation). Under conditions of $k_t > $1.5 GeV/c, which 
translates to $Q^2 \sim 10~{\rm (GeV/c)^2}$ and $\langle r \rangle
\sim 0.1$ fm, observation of these effects can be expected.  
Bertsch et al. \cite{bbgg} proposed that the small  $|q\bar {q}\rangle$ 
component will be filtered by the nucleus. Frankfurt {\em et al.} \cite{fms} 
show that for $k_t > 1.5 GeV/c$ the interaction with the nucleus is 
completely coherent and $ \sigma ( |q\bar {q}\rangle N \rightarrow$di-jets $N$)
is small. This leads to  an $ A^2 $ dependence of the forward amplitude squared.

\section*{Experimental Results}

Fermilab experiment E791 recorded $2\times 10^{10}$ events
from interactions of a 500 GeV/c  $\pi^-$ beam with carbon and
platinum targets. Details of the experiment are given in \cite{e791}.
Only about 10\% of the E791 data was used for the analysis presented here.
The data were analysed by selecting events in which 90\% of the beam
momentum was carried by charged particles. Jets were identifies using the 
JADE jet-finding algorithm \cite{jade}. To insure clean selection of
two-jet events, a minimum $k_t$ of 1.25 GeV/c was required and their relative
azimuthal angle, which for pure di-jets should be $180^{\circ}$ was
required to be within $20^{\circ}$ of this value.

\begin{figure}[h]
\begin{center}
\resizebox{1.0\textwidth}{!}{
 \includegraphics{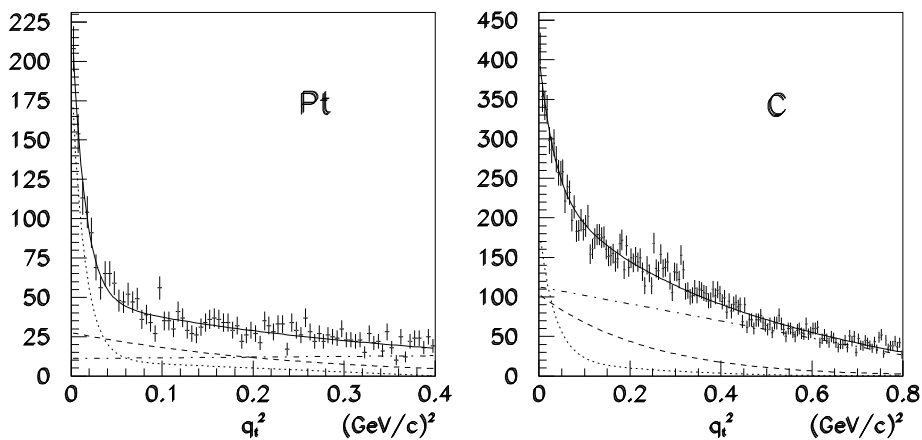}}
\vglue -7.0cm
\caption{$q_t^2$ distributions of di-jets with $1.5 \leq k_t \leq 2.0$
GeV/c for the platinum and carbon targets. The lines are fits of the MC
simulations to the data: coherent dissociation (dotted line), incoherent
dissociation (dashed line), background (dashed-dotted line), and total fit
(solid line).
\label{difr_a}
}\end{center}
\end{figure}

Diffractive di-jets were identified through the $e^{-bq_t^2}$ dependence
of their yield ($q_t^2$ is the square of the transverse momentum transferred
to the nucleus and $b = \frac{<R^2>}{3}$ where R is the nuclear radius).
Figure~\ref{difr_a} shows the $q_t^2$ distributions
of di-jet events from platinum and carbon. The different
slopes in the low $q_t^2$ coherent region reflect the different
nuclear radii. Events in this region come from diffractive dissociation
of the pion. 

\vglue -2.0cm
\subsection*{The Pion Wave Function}
\begin{figure}[h]
\begin{center}
\resizebox{1.0\textwidth}{!}{
 \includegraphics{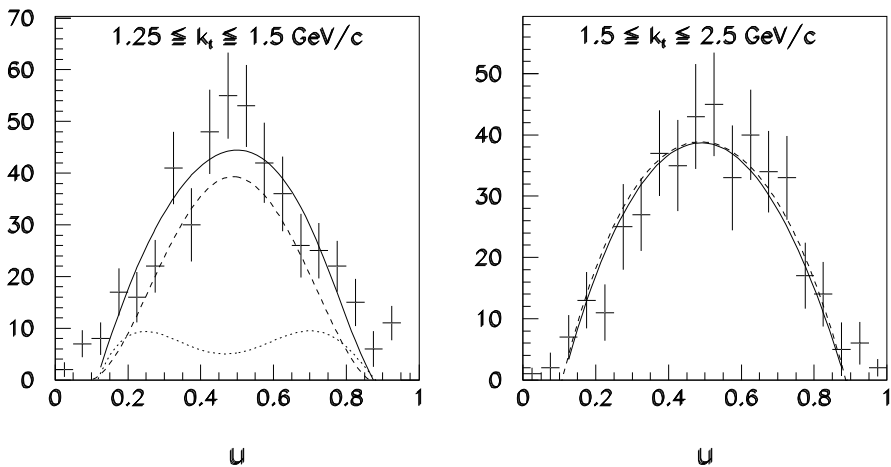}}
\vglue -6.0cm
\caption{The $u$ distribution of diffractive di-jets from
the platinum target for $1.25 \leq k_t \leq 1.5$ GeV/c (left) and for
$1.5 \leq k_t \leq 2.5$ GeV/c (right). The solid line is a fit to a
combination of the asymptotic and CZ wave functions. The dashed line shows
the contribution from the asymptotic function and the dotted line that of
the CZ function.
\label{xdatadif}
}\end{center}
\end{figure}

For measurement of the wave function we used data from
the platinum target as it has a sharp diffractive distribution and low
background. We used events with $q_t^2 ~< 0.015$ GeV/c$^2$. For these events, 
the value of $u$ was computed from the measured longitudinal momentum of each
jet.  A background, estimated from the $u$
distribution for events with larger $q_t^2$ was subtracted.
The analysis was carried out in two windows of $k_t$:
$1.25 ~\rm{GeV/c} ~\leq ~k_t ~\leq ~1.5 ~\rm{GeV/c}$ and
$1.5 ~\rm{GeV/c} ~\leq  ~k_t ~\leq ~2.5 ~\rm{GeV/c}$. 
The resulting $u$ distributions are shown in Fig. \ref{xdatadif}.
In order to get a measure of the correspondence between the experimental
results and the calculated light-cone wave functions, we fit the results
with a linear combination of squares of the two wave functions
(right side of Fig. \ref{fig:x_mc}). This
assumes an incoherent combination of the two wave functions and that the
evolution of the CZ function is slow (as stated in \cite{cz}). 
\begin{table}
\tabcolsep=1.7mm
    \begin{tabular}{|c|c|c|c|c|c|c|c|c|}
\vspace*{-20pt} & & & & & & & & \\ \hline
\vspace*{-10pt} & & & & & & & & \\
$k_t$ (GeV/c) &$a_{as}$ &$\Delta_{a_{as}}^{stat}$ & $\Delta_{a_{as}}^{sys}$ &
 $\Delta_{a_{as}}$ &  $a_{cz}$ & $\Delta_{a_{cz}}^{stat}$ &
$\Delta_{a_{cz}}^{sys}$ & $\Delta_{a_{cz}}$ \\
\vspace*{-10pt} & & & & & & & & \\ \hline
\vspace*{-10pt} & & & & & & & & \\
1.25 - 1.5 &0.64 &$\pm0.12$&+0.07 -0.01& +0.14 -0.12 & 0.36 &$\mp0.12$
&-0.07 +0.01& -0.14 +0.12  \\
1.5 - 2.5 & 1.00  &$\pm0.10$& +0.00 -0.10 & +0.10 -0.14 &  0.00
&$\mp0.10$& -0.00 +0.10 &-0.10 +0.14 \\
\hline
    \end{tabular}
\caption{Asymptotic ($a_{as}$) and CZ ($a_{cz}$)
wave functions contributions to a fit of the data.}
    \label{res}
\end{table}
The results of the fits are given in Fig. \ref{xdatadif} and in Table \ref{res}
in terms of the coefficients $a_{as}$ and $a_{cz}$ representing the
contributions of the asymptotic and CZ functions, respectively. 
The results for the higher $k_t$ window show that
the asymptotic wave function describes the data very well.
Hence, for $k_t > $1.5 GeV/c, which translates 
to $Q^2 \sim 10~{\rm (GeV/c)^2}$, the pQCD approach that 
led to construction of the asymptotic wave function is reasonable.
The distribution in the lower window is consistent with a significant 
contribution from the CZ wave function or may indicate
contributions due to other non-perturbative effects.

\begin{figure}[t]
\begin{center}
\resizebox{1.0\textwidth}{!}{
 \includegraphics{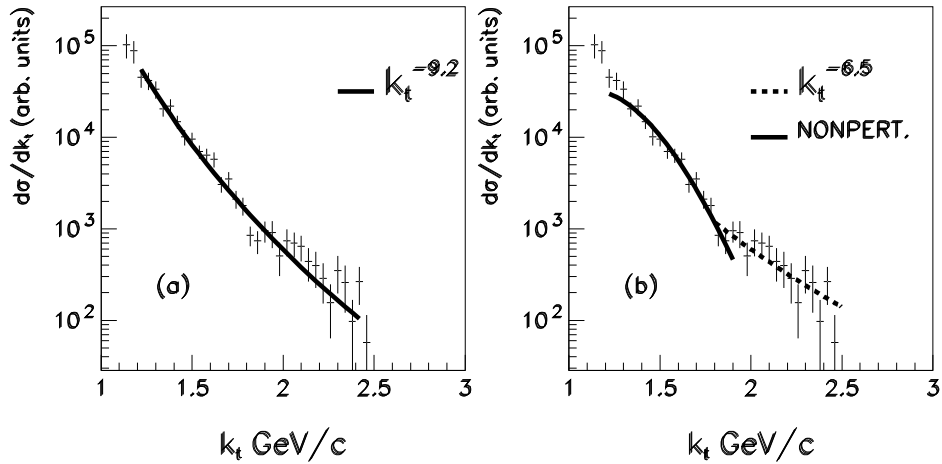}}
\vglue -6.0cm
\caption{Comparison of the $k_t$ distribution of acceptance-corrected data
with fits to
cross section dependence (a) according to a power law, (b) based on a
nonperturbative Gaussian wave function for low $k_t$ and a power
law, as expected from perturbative calculations, for high $k_t$.
\label{split}
}\end{center}
\end{figure}

The $k_t$ dependence of diffractive di-jets is another observable that can
show how well the perturbative calculations describe the data.
As shown in \cite{fms} assuming interaction via two gluon exchange
and $\phi_{as}$ would lead to ${d\sigma\over dk_t} ~\sim ~k_t^{-6}$. 
The results, corrected for experimental acceptance, are shown in
figure \ref{split}(a) fitted by $k_t^n$ for $k_t > 1.25$ GeV
with n = $-9.2 \pm 0.4 (stat) \pm 0.3
(sys)$ and $\chi^2/dof$ = 1.0. This slope is significantly larger than
expected. However, the region above $k_t \sim$ 1.8 GeV/c can be fitted
(Fig. \ref{split}(b)) with n = $-6.5 \pm 2.0$ with $\chi^2/dof$ =
0.8, consistent with the predictions. This would support the evaluation of
the light-cone wave function at large $k_t$ as due to one gluon exchange.
The lower $k_t$ region can be fitted with the non-perturbative Gaussian 
function: $\psi \sim e^{-\beta k_t^2}$ \cite{kro1}, with
$\beta ~= ~1.78 ~\pm ~0.05 (stat) \pm 0.1 (sys)$ and $\chi^2/dof$ = 1.1. 
Model-dependent values in the range of 0.9 - 4.0 were used \cite{kro1}.
These results are consistent with the measurements of the wave function
that indicated noticeable non-perturbative effects up to 
$k_t ~\sim $ 1.5 GeV/c.
\subsection*{The Color Transparency Effect}
\begin{table}[t]
  \begin{center}
    \begin{tabular}{|c|c|c|c|c|c|}
\vspace*{-20pt} & & & & &  \\ \hline
\vspace*{-10pt} & & & & &  \\
$k_t$ bin & $\alpha$ & $\Delta \alpha_{stat}$ &$\Delta \alpha_{sys}$ &
$\Delta \alpha$ & $\alpha$ (CT) \\
GeV/c & & & & & \\
\vspace*{-10pt} & & & & &  \\ \hline
\vspace*{-10pt} & & & & &  \\
1.25 $-$ 1.5 & 1.64     & $\pm$0.05  & +0.04 $-$0.11 &+0.06 $-$0.12
&1.25 \\
1.5 $-$ 2.0 & 1.52     & $\pm$0.09 & $\pm$0.08 &$\pm$0.12  & 1.45 \\
2.0 $-$ 2.5 & 1.55 & $\pm$0.11 & $\pm$0.12 &$\pm$0.16  & 1.60 \\
\hline
    \end{tabular}
\vglue 0.5cm
    \caption{The exponent in $\sigma \propto A^{\alpha}$, experimental
results for coherent dissociation and the Color-Transparency
(CT) predictions.}
\vspace*{5 mm}
    \label{alpha}
  \end{center}
\end{table}


To study the CT effect we measure the A-dependence of the
diffractive di-jet yield. The coherence length is estimated using
$2p_{lab}$ ~= ~1000 GeV/c and  $M_J ~\sim$ 5 $~{\rm GeV/c^2}$. The result
is ~$l_c ~\sim$ ~10 ~fm, larger than the platinum
nuclear  radius. In order to correct for experimental acceptance, we
generate MC simulations of diffractive di-jets using the asymptotic wave 
function and di-jet masses of 4,5, and 6 GeV/c.  
The simulated coherent $q_t^2$ distributions of the di-jets represent the 
nuclear form factors of carbon (R=2.44 fm) and platinum (R=5.27 fm) \cite{rad} 
and the incoherent dissociation is simulated according to the
nucleon radius (R=0.8 fm) \cite{rad} truncated at $q_t^2 < $ 0.015. A 
combination of these simulations was used to fit the data (Fig. \ref{difr_a}).
We derive the numbers of producted di-jet events in the data
for each target in three $k_t$ bins by integrating over the diffractive terms
in the fits. Using the resulting yields and the
known target thicknesses, we determine the ratio of cross sections for
diffractive dissociation on platinum and carbon (the two targets were
subjected to essentially the same beam flux). The exponents $\alpha$ 
are then calculated using the cross section dependence 
$\sigma \propto A^{\alpha}$. The results are listed in Table \ref{alpha} 
and compared with CT theoretical predictions \cite{fms}.
The results are consistent with
those expected from color-transparency calculations and clearly
inconsistent with $\alpha$ values for incoherent scattering
observed in other hadronic interactions.

\section*{The Photon Light-Cone Wave Function}
The photon l.c. wave function can be described in a way similar to
that of the pion except
that it has two major components: the electromagnetic and the hadronic.
Being a gauge field capable of piont-like coupling it has also a point
bare-photon component. Consequently, the photon light-cone wave function
in the pQCD regime can be expanded in terms of Fock states:
\begin{eqnarray}
\psi_{\gamma} = a |\gamma_p \rangle + b |l^+ l^-\rangle + c |l^+ l^-
\gamma \rangle + (other ~e.m.) \nonumber \\
   + d |q\bar {q}\rangle + e |q\bar
{q}g\rangle + (other ~hadronic) + ... .
\end{eqnarray}
where $|\gamma_p \rangle$ describes the point bare-photon and
$|l^+ l^-\rangle$ stands for $|e^+ e^-\rangle,~|\mu^+ \mu^-\rangle$ etc.
The photon can be described as containing $|q\bar{q}\rangle$ components
even in the non-pQCD regime of the vector-meson dominance model \cite{fey},
in particular the $\rho, ~\omega, \phi$ and $J/\psi$ mesons. There is also
a $|q\bar{q}\rangle$ component with large relative transverse momentum
that is connected to the point-like (direct) photon. The wave function
of the photon is very rich: it can be studied for real photons, for virtual
photons of various virtualities, for transverse and longitudinal photons
and the hadronic component may be decomposed
according to the quarks flavor. Strange quarks will reflect the strange
content of the photon and charmed quarks will reflect the smaller sized
charmed components. 

\begin{figure}
\begin{center}
\resizebox{1.0\textwidth}{!}{
 \includegraphics{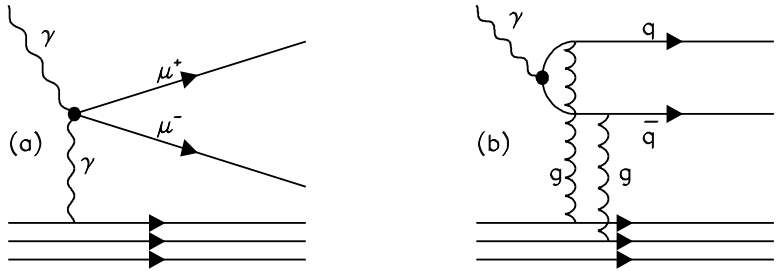}}
\vglue -8.0cm
\caption{Diagrams for photon dissociation}
\label{diag2}
\end{center}
\end{figure}

 The point-bare photon does not have internal structure, it can only
Compton-scatter and will not be studied here. The other electromagnetic
Fock states begin with $|l^+l^-\rangle$ and continue to more complex
systems. The interaction of these Fock states with the target is expected 
to be purely electromagnetic, as shown in figure \ref{diag2}(a). These
were studied extensively through measurements of Bethe-Heitler 
leptoproduction \cite{bh} and of $F_{2,QED}^{\gamma}$
\cite{nis}. Exclusive measurements of the $|l^+l^-\rangle$ wave function
will complement these inclusive measurements, may provide more tests of
QED and will help define the analysis tools for the more complex hadronic
structure.

\begin{figure}
\begin{center}
\resizebox{0.65\textwidth}{!}{
 \includegraphics{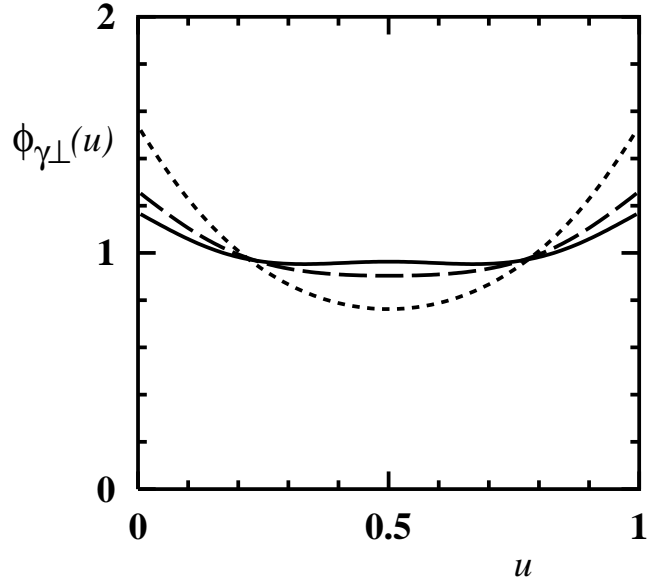}}
\vglue -1.0cm
\caption{The Photon wave functions, Petrov et al. Real photons (Solid
line), virtual photons Q$^2$ = 250 MeV$^2$ (dashed line), Q$^2$ = 500
MeV$^2$ (dotted line).}
\label{phi_we}
\end{center}
\end{figure}

The hadronic components interact according to the diagram of Fig.
\ref{diag2}(b). The distribution amplitudes for these components, 
integrated over $k_t$ were calculated by
Petrov et al. \cite{weiss} using the instanton model and are shown in
figure \ref{phi_we}. Balitsky et al. \cite{bal} predict for real photons a
distribution amplitude identical to the pion asymptotic function
\cite{lb,er,bbgg} (see Fig. \ref{fig:x_mc}).

While a variety of cross sections, form factors etc. depend on the
light-cone wave function they are usually not sensitive to its structure.
This is because it normally enters the calculations in an integrated form.
Such are also the photon inclusive structure functions $F_2^{\gamma}$ 
\cite{nis}. While this is an advantage when one needs calculations that do
not depend on the internal structure of the photon, it is a disadvantage 
if we want to understand it. A differential measurement which will
be sensitive to the $u$ and $k_t$ dependence of $\phi(u,Q^2)$ will test the
most fundamental description of the photon internal structure. Such
measurements will make it possible to compare with the theoretical
predictions and determine the regime of their validity. 

The experimental program is presently in preparation 
and is based on diffractive dissociation of a real or virtual photon
into di-leptons or di-jets. The electromagnetic component will be studied
through measurements of the $u$ distribution for pure diffractive
$\mu^+\mu^-$ ($e^+e^-$) elastic photoproduction and the $k_t$ 
distribution of these events. Similar measurements of diffractive
hadronic di-jets will be used to study the hadronic component.

I would like to acknowledge the efforts of the E791 collaboration, of
which I am a member, for the data presented in this work and 
the data analysis performed by my graduate
student R. Weiss-Babai. 

\end{document}